\documentclass[%
reprint,
amsmath,amssymb,
aps,
prb,
]{revtex4-1}
\usepackage{float}

\usepackage{hyperref}

\usepackage{graphics}
\usepackage{graphicx}% Include figure files
\usepackage{dcolumn}% Align table columns on decimal point
\usepackage{bm}% bold math
\begin{document}

\title{Spectra of the Dissipative Spin Chain}% Force line breaks with \\

\author{Jian Wang}
\author{Sudip Chakravarty}%
\affiliation{%
 Department of Physics and Astronomy, University of California Los Angeles, California 90095\\
 Mani L Bhaumik Institute for Theoretical Physics
}%

\date{\today}% It is always \today, today,

\begin{abstract}
This paper generalizes the (0+1)-dimensional spin-boson problem to the corresponding (1+1)-dimensional version. Monte Carlo simulation is used to find the phase diagram and imaginary time correlation function. The real frequency spectrum is recovered by the newly developed P\'ade regression analytic continuation method. 
We find that, as dissipation strength $\alpha$ is increased, the sharp quasi-particle spectrum is broadened and the peak frequency is lower.  According to the behavior of the  low frequency spectrum, we classify the dynamical phase into three different regions: weakly damped, linear $k$-edge, and strongly damped.

\end{abstract}

\pacs{Valid PACS appear here}% PACS, the Physics and Astronomy
\maketitle

%\tableofcontents

\section{Introduction}
Dissipation plays an important rule in quantum phase transitions  \cite{Chakravarty1984,RevModPhys.59.1,Sudip1995,Weiss2012,PhysRevB.98.241108, PhysRevA.95.042133,decoherence2004,decoherence2018,JosephsonJunction2018,Cai2018}. There can be localization-delocalization transitions and coherence-decoherence transitions as the  dissipative strength is tuned. Dissipative dynamics is also the bottleneck to build a reliable quantum computer. \cite{1812.04471,PhysRevA.98.052109}  However, exactly solvable dissipative quantum systems are few and far between and often numerical approaches are needed, However, extracting reliable real time dynamics from numerical simulation in the imaginary time simulation is difficult. Ironically, it is the real time results that are mostly relevant to experiments.

In this work, we are going to extend the (0+1) dimensional \cite{Dekker1987,PhysRevB.58.1862} spin-boson system to (1+1) dimension. It is a transverse Ising chain, with each spin coupled to a Ohmic bosonic heat bath.

We use Monte Carlo method \cite{Wolff1989,
Wang1987,
10.1007/978-3-642-59689-6_7, doi:10.1063/1.5043096} to explore the system and generate imaginary time spin-spin correlations \cite{PhysRevLett.94.047201,PhysRevB.81.104302,XY2005}. For  analytic continuation to the real time, we  use our newly developed Pad\'e Regression method \cite{PRM} to get the real time dynamical spectra.

In the limit of no dissipation, the real frequency spectrum can be exactly solved via Jordan-Wigner transformation \cite{Pfeuty1970,Young1996,myself_1}. Hence our quantum Monte Carlo and the analytic continuation methodology can be checked to some extent by comparing  with the exact results in the case of no dissipation Fig.~\ref{fig:swk00}. In Sec. II we define the model and describe the Monte Carlo simulation in III. In Sec. IV we discuss the results and the conclusions are discussed in V.

\section{spin chain in a dissipative bath }

The model has 3 parts: \(H_S\) is the transverse field Ising chain, \(H_B\) is the dissipative bosonic bath, \(H_I\) is the coupling of the Ising chain with the bath. The influence of environment to the \(i\)-th spin in the Ising chain can be completely describe by the correlation \( J_i(\omega)=\sum_k c_{i,k}^2 \delta(\omega-\omega_{i,k} ) \). By assuming Ohmic bath, we are assuming that the correlation takes the linear form at low frequency: \(J(\omega) = 2\pi \alpha \omega e^{-\omega/\omega_0} \), where \(\omega_0\) is some high energy cut off, it doesn't affect the lower energy physics.

\begin{align}
	H & = H_S+H_I+H_B   \label{eq:sb4} \\
	H_S &=-\Delta \sum_{i=1}^L \sigma_i^x - J \sum_{i=1}^L  \sigma_i^z \sigma_{i+1}^z \nonumber \\
	H_I &= \sum_{i=1}^{L} \sum_{k=1}^{N}c_{i,k}{\bigg(a^\dagger_{i,k} + a_{i,k}\bigg)} \sigma_i^z \nonumber \\
	H_B &= \sum_{i=1}^{L}\sum_{k=1}^{N} \omega_{i,k} (a^\dagger_{i,k} a_{i,k} +\frac{1}{2})  \nonumber
\end{align}

Path integral formalism is carried out to map the quantum Hamiltonian into classical action \cite{Chakravarty1984}. The dissipative Bosonic heat bath is traced out, leaving a $1/r^{2}$  longer range interaction in imaginary time ($\propto \tau $), \(\alpha\) becomes \(A_0\).  \(\sin^2\) is for the periodic boundary condition. \cite{  Luijten2001}

\begin{table}[h]
\caption{\label{tab:example}classical-quantum mapping}
\begin{ruledtabular}
\begin{tabular}{lll}
quantum & classical & relation\\
$L$ & $N_1$ & $L=N_1$\\
$\beta$ & $N_0$ & $\beta=N_0$\\
$J$ & $K_1$ & $J=K_1$\\
$\Delta $ & $K_0$ & $\tanh(\Delta)=\exp(-2K_0)
$
\end{tabular}
\end{ruledtabular}
\end{table}

	\begin{eqnarray}
			S_{\text{ classical action}}=-K_1 \sum \sum s_{i,\tau} s_{i+1,\tau} \nonumber \\
			-K_0\sum \sum s_{i,\tau} s_{i,\tau+1} \nonumber \\ -\frac{\alpha}{2} \sum \sum_{\tau<\tau'} s_{i,\tau} s_{i,\tau'} \big(\frac{\pi}{N_{\tau}}\big)^2 \frac{1}{ \sin ^2 \big( \frac{\pi}{N_\tau}  |\tau-\tau'| \big)}
		\end{eqnarray}

\section{Monte Carlo method}		
The Monte Carlo simulation is carried out on system sizes \(N_0\times N_1=128 \times 64\)  with Wolff clustering updating algorithm. The total updating steps are $[\text{Jump}]\times 2^{26}$. Here we update every $[\text{Jump}]$ steps to keep the samples as uncorrelated as possible.  In order to increase the acceptance rate of long range interaction in the imaginary time, \(N_0\),  direction, cumulative probability method is applied \cite{10.1007/978-3-642-59689-6_7}.  We ran on a single CPU core for two weeks; the relative error for the $I[\omega_n,k]$ (see below) is less than 0.1\%.

\subsection{Spin-spin correlation}

\subsubsection{The standard method}
Given 2D Ising spin $s[\tau,x]=\pm 1$  on a discrete lattice with periodic boundary condition, where $\tau\in \{0,1,2,\cdots N_0-1 \}$ is in the imaginary time direction and $x\in \{0,1,2,\cdots N_1-1 \} $ is in the spatial direction, our goal is to calculate spin-spin correlation function $c[\tau,x] =\langle s[0,0]s[\tau,x] \rangle $.  Here $\langle \cdots \rangle $ is the Monte Carlo  average.

Since our problem is  translational invariant. We also have   $c[\tau,x] =\langle s[1,1]s[1+\tau,1+x] \rangle = \langle s[1,2]s[1+\tau,2+x] \rangle = \cdots =  \langle s[\tau_0,x_0]s[\tau_0+\tau,x_0+x] \rangle $   for any initial site $\tau_0,x_0$. Therefore we can write the correlation function as:
\begin{equation}
    c[\tau,x]= \bigg\langle \frac{1}{N_0 N_1} \sum_{\tau_0=0}^{N_0-1}\sum_{x_0=0}^{N_1-1} s[\tau_0,x_0]s[\tau_0+\tau,x_0+x] \bigg\rangle
    \label{eq:c_tau_x}
\end{equation}
We need to perform $N_0N_1$ multiplications to get one value of $c[\tau,x]$. There are $N_0N_1$ values of $c[\tau,x]$ for each index $[\tau,x]$. Therefore, to get a 2D correlation function $c[\tau,x]$, we need $O( (N_0 N_1)^2 M) $ total multiplications.  Where $M$ is the Monte Carlo updating steps.

Then we can perform a 2D discrete Fourier transform on $c[\tau,x]$  to get the $I[\omega_n,k]$
\begin{equation}
    I[\omega_n,k]  = \frac{1}{\sqrt{N_0 N_1}} \sum_{\tau=0}^{N_0-1}\sum_{x=0}^{N_1-1} e^{  i (\tau\omega_n+xk)} c[\tau,x]
    \label{eq:I_w_k}
\end{equation}
If we make the analytic continuation from Matsubara frequency $i\omega_n$ to real frequency $\omega$, the function $I[\omega_n,k]$ becomes $S[\omega,k]$. It is the dynamical structure factor of the quantum spin system.
 \subsubsection{A faster method}
The convolution theorem and fast Fourier transform can make the above calculation faster. The acceleration is from $O((N_0N_1)^2M)$ to  
$O(N_0N_1\log(N_0N_1)M)$. The equation is given by

\begin{equation}
    I[\omega_n,k] =\bigg\langle  \bigg|  \tilde{s}[\omega_n,k]\bigg|^2\bigg\rangle
    \label{eq:FFTsquared}
\end{equation}
where $\tilde{s}[\omega_n,k]$ is the 2D discrete Fourier transformation of the Ising spin field $s[\tau,x]$
\begin{equation}
    \tilde{s}[\omega_n,k]=\frac{1}{\sqrt{N_0 N_1}} \sum_{\tau=0}^{N_0-1}\sum_{x=0}^{N_1-1} e^{  i (\tau\omega_n+xk)} s[\tau,x] \label{eq:spinFFT}
\end{equation}
We have used  the fact that the order of the Fourier transform and the summation  can be exchanged due to linearity. Equation~(\ref{eq:spinFFT}) and \ref{eq:FFTsquared}) will give the same $I[\omega_n,k]$ as Eqs.~(\ref{eq:c_tau_x})  and  (\ref{eq:I_w_k}), but with a logarithmic acceleration.

\subsection{Analytic continuation}

To begin, we have a classical system of size \(N_0\times N_1=128\times64\). Consider  the  correlation \( C[\tau,x]= \langle s[\tau_0,x_0] s[\tau_0+\tau,x_0+x]  \rangle \) and  perform a 2D discrete Fourier transformation on \(C[\tau,x]\), to get \( I[\omega_n,k] \), which is also the quantum \(G(i\omega_n,k)\). The values of \(\omega_n, k\) run through discrete points in the Brillouin zone. Where \(\Omega=\frac{2\pi}{\beta}=\frac{2\pi}{N_0}\) is the Matsubara frequency interval.
\begin{eqnarray}
G(i\omega_n,k)\equiv  I[\omega_n,k] \\
\omega_n  = 0, \Omega, 2\Omega, \cdots ,(N_0-1)\Omega \nonumber \\
k = 0,  \frac{2\pi}{N_1}, 2\frac{2\pi}{N_1},\cdots , (N_1-1)\frac{2\pi}{N_1}
\end{eqnarray}

\begin{eqnarray}
\label{eq:analyticContinuation}
G(i\omega_n,k) \rightarrow G(\omega+i0^+,k) \rightarrow S(\omega,k) 
\end{eqnarray}

The analytic continuation Eq.~(\ref{eq:analyticContinuation}) is done for each fixed \(k\) value, using our newly developed P\'ade regression method \cite{PRM}. The P\'ade regression assumes the analytic function $G(z)$ takes the specific form of a rational function  $\frac{P_L(z)}{P_M(z)}=\frac{a_0+a_1 z +\cdots + a_L z^L}{1+b_1 z+ \cdots + b_M z^M}$.
The  polynomial in the numerator is of degree is $L$ and the denominator  is of degree $M$. Therefore there are $L+M+1$ parameters to be determined.
Given $N$ Matsubara points, there are $N$ fitting equations  $G(z_n= i\omega_n)=u_n \quad (n=1,2,\cdots,N)$.
We then modify the problem  to  a linear regression problem: given $\mathbf{X}$ and $\mathbf{y}$ find the $\beta$ that minimizes $|| \mathbf{X} \beta -\mathbf{y}||^2 $.  Here the explicit form of $\mathbf{X}_{N \times(L+M+1) } \mathbf{\beta}_{(L+M+1)} = \mathbf{y}_N$  is in Eq.~(\ref{eq:linearRegressionMatrix})

\begin{widetext}
\begin{eqnarray}\hspace*{-2cm} 
\label{eq:linearRegressionMatrix}
\left(
\begin{array}{ccc|cccc}
-u_1 z_1^1 & -u_1 z_1^2 & \hdots &  z_1^0  & z_1^1 & z_1^2 & \hdots \\
-u_2 z_2^1 & -u_2 z_2^2 & \hdots &  z_2^0  & z_2^1 & z_2^2 & \hdots \\
\vdots & \vdots &\vdots &\vdots & \vdots &\vdots & \vdots \\

\vdots & \vdots &\vdots &\vdots & \vdots &\vdots & \vdots \\
\vdots & \vdots &\vdots &\vdots & \vdots &\vdots & \vdots \\
\vdots & \vdots &\vdots &\vdots & \vdots &\vdots & \vdots \\
\vdots & \vdots &\vdots &\vdots & \vdots &\vdots & \vdots \\
-u_N z_N^1 & -u_N z_N^2 & \hdots &  z_N^0  & z_N^1 & z_N^2 & \hdots \\
\end{array} \right)
	\begin{pmatrix}
b_1 \\
b_2 \\
\vdots \\
\vdots \\
\hline
a_0 \\
a_1 \\
a_2 \\

\vdots

\end{pmatrix}=
	\begin{pmatrix}
		u_1 \\
		u_2 \\
		\vdots \\
		\vdots \\
		\vdots \\
		\vdots \\
		u_{N-1} \\
		u_N
	\end{pmatrix}
\end{eqnarray}

\end{widetext}
Starting from this standard linear regression problem, we can apply Bayesian inference to choose the optimal $L$ and $M$ or use bootstrapping to estimate the error.
\section{Result}

\subsection{Calibration}

Let's first  look at the case without dissipation. This is just the transverse field Ising model; the exact spectrum is \( \epsilon(k)= \sqrt{\Delta^2+J^2 - 2 \Delta J \cos(k) } \). Therefore we can use the exact result to verify our Monte Carlo plus analytic  continuation approach. The classical-quantum mapping, will map \(K_0=0.136, K_1=0.2, N_0=128, N_1=64\) to the quantum parameter \(\Delta=1,J=0.2,\beta=128,L=64\).

\begin{figure}[H]
 	\centering\includegraphics[width=0.9\linewidth]{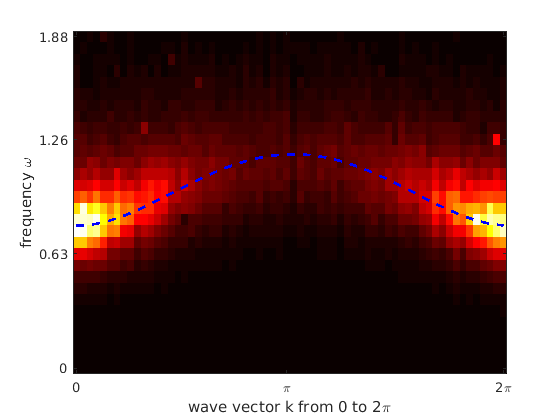}
 	\caption{$S(\omega,k)$ of transverse field Ising chain  $\Delta=1.0, J=0.2$ with no dissipation, $\alpha=0$. The recovered spectrum is compared with the exact result (the dashed blue curve).}
 	\label{fig:swk00}
 \end{figure}

\begin{figure}[H]
 	\centering\includegraphics[width=0.9\linewidth]{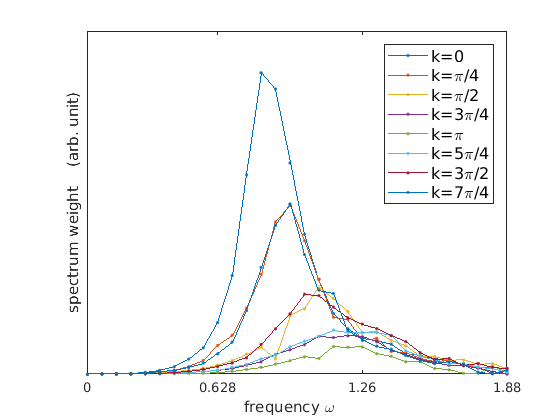}
 	\caption{Transverse field Ising chain $\Delta=1.0, J=0.2$ with no dissipation $\alpha=0$.  Each curve is $S(\omega,k)$ with fixed $k$ value.}
 	\label{fig:swk00cut}
 \end{figure}

Fig \ref{fig:swk00cut} is the \(S(\omega,k) \) result for each individual \(k\). Lower momenta  have always higher spectral weight. We can also see the symmetry of the spectrum, \(S(\omega,k) \) and  \(S(\omega,2\pi-k) \)  have the same  shape: Fig \ref{fig:swk00} is the color  version. The blue dashed line is the exact  spectrum \( \omega(k)=\sqrt{1+0.2^2 - 2\times 0.2 \cos(k)} \), we can see that the exact result and the analytic continuation  agree reasonably well.

The broadening is due to two reasons (1) finite size (classical \(N_0=128\)) or the finite temperature effects (quantum \(T=1/\beta=1/N_0\)) ; (2) our current Monte Carlo imaginary time correlation function  has 5 significant digits (relative error  $10^{-5}$), which is still a large error.

\subsection{Spectrum with dissipation}

\begin{figure}[H]
 	\centering\includegraphics[width=0.9\linewidth]{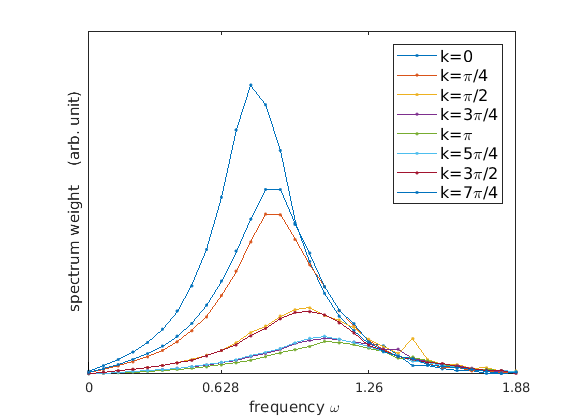}
 	\caption{ Transverse field Ising chain $\Delta=1.0, J=0.2$ with dissipation $\alpha=0.05$.  Each curve is $S(\omega,k)$ with fixed $k$ value.} 
 	\label{fig:swk005cut}
 \end{figure}

\begin{figure}[H]
 	\centering\includegraphics[width=0.9\linewidth]{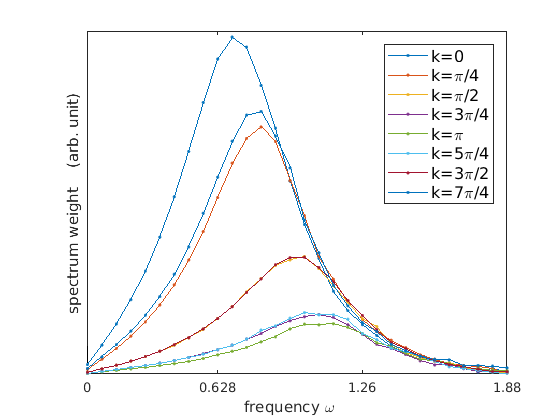}
 	\caption{Transverse field Ising chain $\Delta=1.0, J=0.2$ with dissipation $\alpha=0.1$.  Each curve is $S(\omega,k)$ with fixed $k$ value.}
 	 	\label{fig:swk01cut}
 \end{figure}

\begin{figure}[H]
 	\centering\includegraphics[width=0.9\linewidth]{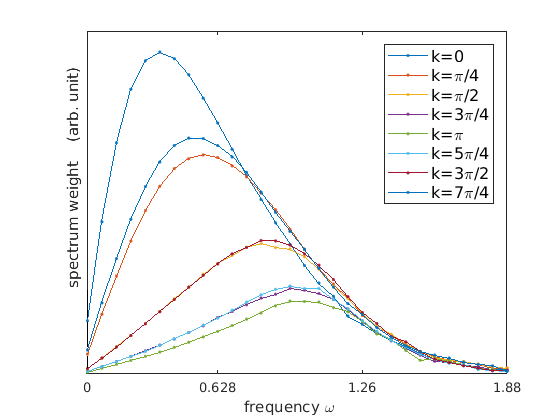}
 	\caption{Transverse field Ising chain $\Delta=1.0, J=0.2$ with dissipation $\alpha=0.2$.  Each curve is $S(\omega,k)$ with fixed $k$ value.}
 	 	\label{fig:swk02cut}
 \end{figure}

\begin{figure}[H]
 	\centering\includegraphics[width=0.9\linewidth]{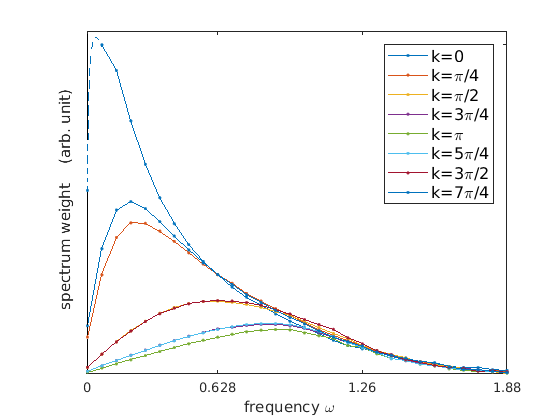}
 	\caption{ Transverse field Ising chain $\Delta=1.0, J=0.2$ with dissipation $\alpha=0.3$.  Each curve is $S(\omega,k)$ with fixed $k$ value. The $k=0$ curve changes violently near zero frequency, we use dashed line to interpolate.}
 	 	\label{fig:swk03cut}
 \end{figure}

\begin{figure}[H]
 	\centering\includegraphics[width=0.9\linewidth]{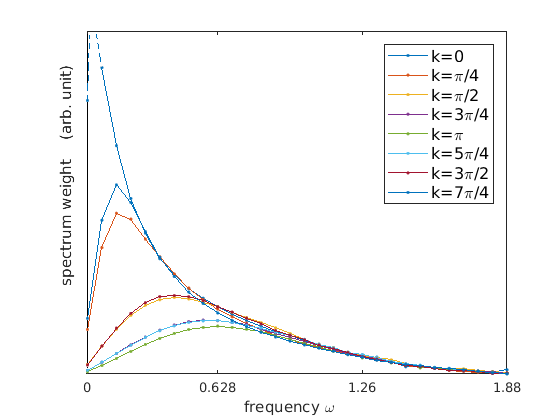}
 	\caption{ Transverse field Ising chain $\Delta=1.0, J=0.2$ with dissipation $\alpha=0.5$.  Each curve is $S(\omega,k)$ with fixed $k$ value. The $k=0$ curve changes violently near zero frequency, the peak will be out of the graph. We use dashed line to interpolate. }
 	 	\label{fig:swk05cut}
 \end{figure}

\begin{figure}[H]
 	\centering\includegraphics[width=0.9\linewidth]{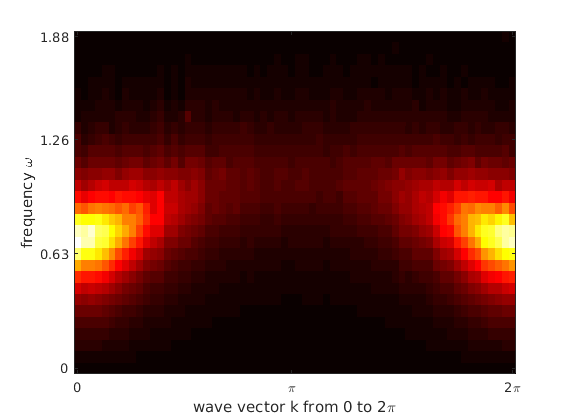}
 	\caption{ $S(\omega,k)$ of transverse field Ising chain  $\Delta=1.0, J=0.2$ with dissipation $\alpha=0.05$.  }
 	\label{fig:swk005}
 \end{figure}

\begin{figure}[H]
 	\centering\includegraphics[width=0.9\linewidth]{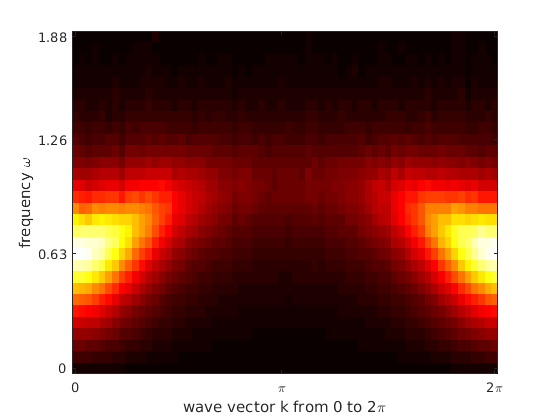}
 	\caption{ $S(\omega,k)$ of transverse field Ising chain  $\Delta=1.0, J=0.2$ with dissipation $\alpha=0.1$.  }
 	\label{fig:swk01}
 \end{figure}

\begin{figure}[H]
 	\centering\includegraphics[width=0.9\linewidth]{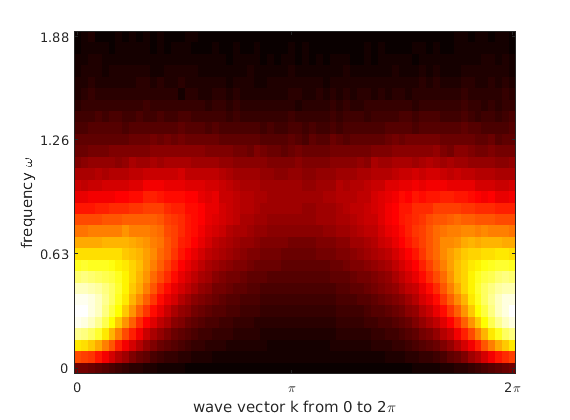}
 	\caption{ $S(\omega,k)$ of transverse field Ising chain  $\Delta=1.0, J=0.2$ with dissipation $\alpha=0.2$. }
 	 	\label{fig:swk02}
 \end{figure}

\begin{figure}[H]
 	\centering\includegraphics[width=0.9\linewidth]{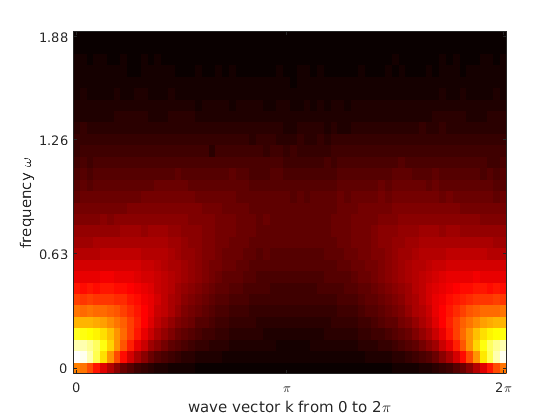}
 	\caption{ $S(\omega,k)$ of transverse field Ising chain  $\Delta=1.0, J=0.2$ with dissipation $\alpha=0.3$. }
 	 	\label{fig:swk03}
 \end{figure}

\begin{figure}[H]
 	\centering\includegraphics[width=0.9\linewidth]{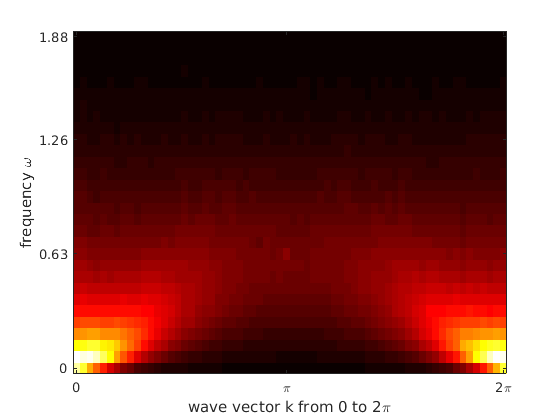}
 	\caption{ $S(\omega,k)$ of transverse field Ising chain  $\Delta=1.0, J=0.2$ with dissipation $\alpha=0.5$.  }
 	 	\label{fig:swk05}
 \end{figure}

 We  turn on the dissipative strengths to be \(\alpha=0.05, 0.1,0.2,0.3, 0.5\). Fig.~(\ref{fig:swk005cut} \ref{fig:swk01cut} \ref{fig:swk02cut}  \ref{fig:swk03cut} \ref{fig:swk05cut} ) are the spectral plots for individual \(k\).  Fig.~( \ref{fig:swk005} \ref{fig:swk01} \ref{fig:swk02}  \ref{fig:swk03} \ref{fig:swk05}) are the corresponding density plots of \(S(\omega,k)\).  From these results, we can see that as the dissipation strength is increased, the energy peak is shifted down.  The energy distributions also get broadened, implying shorter life time of the quasi-particle excitation.

 The energy gap is more subtle. Only in the non-dissipative system, can we observe a clean energy gap. As the dissipation is turned on a little bit, it forms a pseudo-gap, and closes softly.  At low energies \(S(\omega) \propto \omega^\delta \), we can classify the gap closing into three cases: \(\delta>1\) soft closing, \(\delta=1\) linear closing, \(\delta<1\) hard closing.  The low energy exponent \(\delta=\delta(\alpha,k) \)  is a function of dissipation strength \(\alpha\), and momentum \(k\).

For \(\alpha=0.1\), see Fig \ref{fig:swk01cut}. The spectral curve is convex at low energy for all momentum.  \(\delta(0.1, k)<1\)  
For \(\alpha=0.2\), see Fig \ref{fig:swk02cut}. It's very interesting. At low momentum, the  spectrum is convex \(\delta>1\), while at high momentum, the spectrum is concave \(\delta<1\). And there exist a special momentum \(k_c\) such that  the dispersion is linear  \(\delta(\alpha, k_c)=1\), which divides the convex and concave regions. (in the \(\alpha=0.2\) case, it is  \(k_c\approx \frac{\pi}{2}, \pi\))
For \(\alpha=0.3\), see Fig \ref{fig:swk03cut}. The spectrum shifts to low frequency and the gap is closing. The low energy shape is concave (\(\delta <1 \)) for all momenta.

\subsection{Three dynamical phases}

As dissipation is turned on, the low momentum spectrum  gets damped faster than the high momentum, in terms of the \( \delta \) value. Therefore we can classify the system into three different regions:
\begin{enumerate}
    \item Weakly damped region
    \item Linear $k$-edge region
    \item Strongly damped region
\end{enumerate}

\begin{figure}[H]
 	\centering\includegraphics[width=1\linewidth]{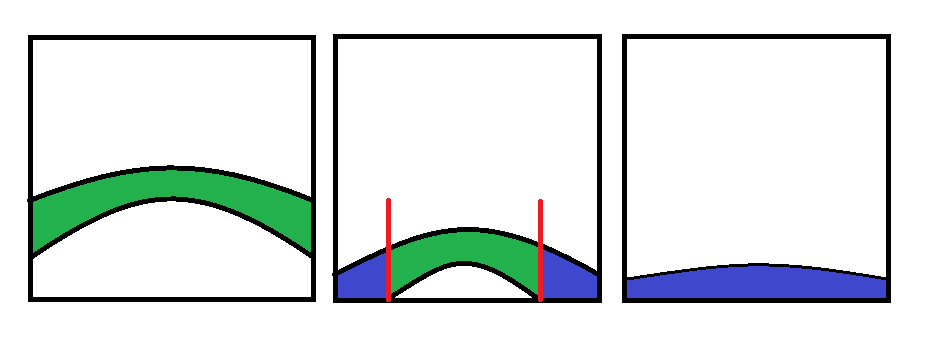}
 	\caption{ A schematic showing three dynamical phases (from left to right: weak damped $\alpha=0.1$, linear k-edge $\alpha=0.2$, strong damped $\alpha=0.3$ ). The horizontal axis is $k$, the vertical axis is $\omega$, same as Fig.~(\ref{fig:swk01},\ref{fig:swk02},\ref{fig:swk03}). The band is colored green if \(\delta(k) > 1\) , blue if \(\delta(k) < 1\)  and red if \(\delta(k) = 1\).  }
 	 	\label{fig:threeCases}
 \end{figure}

\begin{figure}[htb]
 	\centering\includegraphics[width=1.2\linewidth]{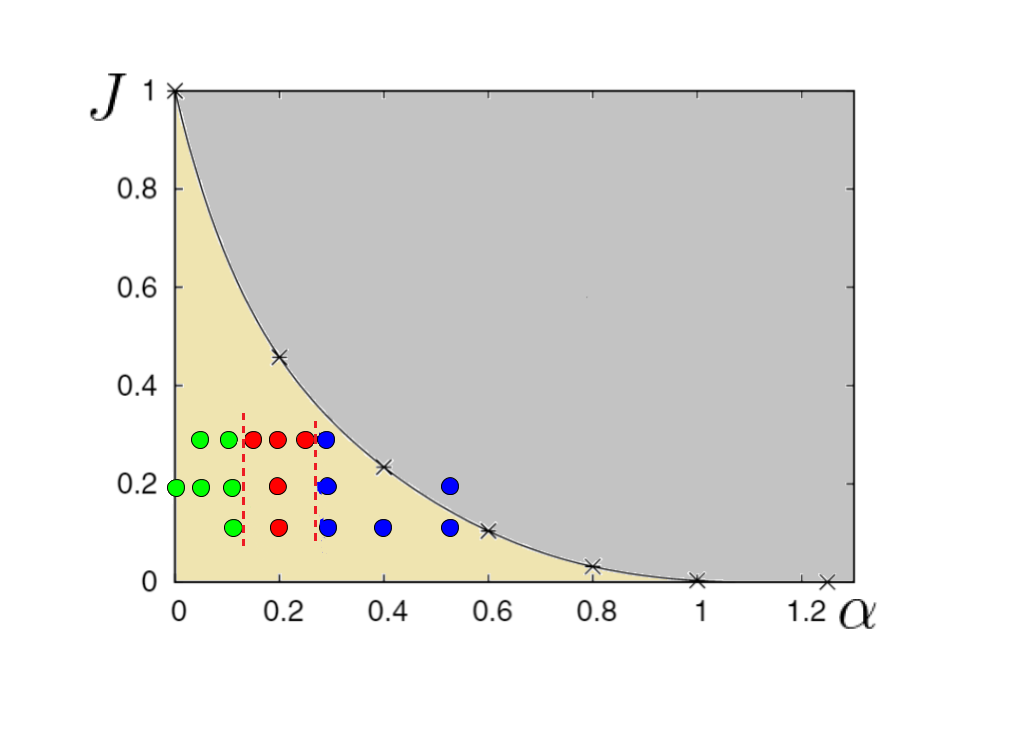}
 	\caption{ Phase diagram of dissipative Ising chain. Vertical axis \(J\) is the nearest neighbour \(\sigma^z_i \sigma^z_{i+1}\)coupling, horizontal axis \(\alpha\) is the dissipation strength, transverse field is set to \(\Delta=1\) or \(K_0=0.136\).  The yellow and grey regions are the disordered and ordered magnetic phases from the imaginary time simulation \cite{PhysRevLett.94.047201}.  Green, red, and blue dots represent weakly damped, linear $k$-edge, and strongly damped regions respectively. }
 	 	\label{fig:phaseDiagram}
 \end{figure}

In Fig \ref{fig:threeCases}, the schematics of these three regions are plotted. Fig \ref{fig:phaseDiagram} is the  phase diagram. The light yellow and grey region correspond to the magnetically disordered and ordered phases in the imaginary time simulation. Green, red, blue dots correspond to the three dynamical phases of the real time spectra. 

In the limit of zero dissipation, it is the transverse field Ising model, which is an integrable system. For each $k$  the excitation has infinite life time.
In the limit of large dissipation, the Hamiltonian is dominated by the environmental noise term.   The quasi-particles will decay faster than its energy time scale. In the intermediate dissipation range, low momentum will not have quasi-particle excitation, while high momentum will. The critical damping edge momentum \(k_c\), is given by \(S(\alpha,\omega ,k_c) \propto \omega \). \\

\section{Conclusion}
To summarize, we have used extensive quantum Monte Carlo simulation, plus the rational function (Pad\'e) regression method to recover the spectra of the dissipative Ising chain. As the dissipation strength is increased, the spectral speak is broadened and lowered in energy. Quasi-particle picture $S(\omega,k)= \delta(\omega-\omega(k))$ does not hold; $\frac{1}{\omega-\omega'(k)-i \omega''(k)} $ is generalized to an arbitrary rational function.   According to lower energy exponent of $S(\omega,k) \sim \omega^{\delta(k)} $ three dynamical regions are introduced to understand the role of dissipation.

\begin{acknowledgments}
This work used computational and storage services associated with the Hoffman2 Shared Cluster provided by UCLA Institute for Digital Research and Education's Research Technology Group. The research was supported in part by funds from David S. Saxon Presidential Term Chair.
\end{acknowledgments}

%\bibliography{citations}


\end{document}